\documentclass[reprint,superscriptaddress,
amsmath,amssymb,
aps,
pra
]{revtex4-1}

\usepackage{qip}
\usepackage{graphicx}
\usepackage[colorlinks=true,citecolor=blue,urlcolor=black]{hyperref}
\usepackage{amsfonts}
\usepackage{amsmath,amssymb}
\usepackage{dsfont}
\usepackage{euscript}
\usepackage{float}
\usepackage{amsthm}
\usepackage{mathalfa}
\usepackage{bbm}
\usepackage{ragged2e}
\usepackage{color}
\usepackage{xcolor}
\usepackage{pgf}

\usepackage{color,soul}
\usepackage{tikz}
\usetikzlibrary{backgrounds}
\def\be{\begin{equation}}
\def\ee{\end{equation}}

\newcommand{\Co}{\mathcal{C}}

\newcommand{\As}{\mathsf{a}}
\newcommand{\Bs}{\mathsf{b}}
\newcommand{\Es}{\mathsf{e}}
\newcommand{\pass}{\mathtt{pass}}
\newcommand{\fail}{\mathtt{fail}}

\newcommand{\U}{\mathcal{U}}
\newcommand{\K}{\mathcal{K}}
\newcommand{\T}{\mathcal{T}}
\newcommand{\X}{\mathbb{X}}
\newcommand{\Y}{\mathbb{Y}}
\newcommand{\Z}{\mathbb{Z}}
\newcommand{\real}[1]{\text{Re} \left[ #1 \right]}

\newcommand{\ebit}{e_{\text{bit}}}
\newcommand{\eph}{e_{\text{ph}}}
\newcommand{\vac}{\text{vac}}

\begin{document}

\title{Versatile security analysis of measurement-device-independent quantum key distribution}

\author{Ignatius William Primaatmaja}\email{ign.william@gmail.com}
\affiliation{Centre for Quantum Technologies, National University of Singapore, Singapore}
\author{Emilien Lavie}
\affiliation{Centre for Quantum Technologies, National University of Singapore, Singapore}
\affiliation{T\'el\'ecom ParisTech, LTCI, Paris, France}
\affiliation{Department of Electrical \& Computer Engineering, National University of Singapore, Singapore }
\author{Koon Tong Goh}
\affiliation{Department of Electrical \& Computer Engineering, National University of Singapore, Singapore }
\author{Chao Wang}
\affiliation{Department of Electrical \& Computer Engineering, National University of Singapore, Singapore }
\author{Charles Ci Wen \surname{Lim}}
\affiliation{Centre for Quantum Technologies, National University of Singapore, Singapore}
\affiliation{Department of Electrical \& Computer Engineering, National University of Singapore, Singapore }

\begin{abstract}
Measurement-device-independent quantum key distribution (MDI-QKD) is the only known QKD scheme that can completely overcome the problem of detection side-channel attacks. Yet, despite its practical importance, there is no standard approach towards proving the security of MDI-QKD. Here, we present a simple numerical method that can efficiently compute almost-tight security bounds for any discretely modulated MDI-QKD protocol. To demonstrate the broad utility of our method, we use it to analyze the security of coherent-state MDI-QKD, decoy-state MDI-QKD with leaky sources, and a variant of twin-field QKD called phase-matching QKD. In all of the numerical simulations (using realistic detection models) we find that our method gives significantly higher secret key rates than those obtained with current security proof techniques. Interestingly, we also find that phase-matching QKD using only two coherent test states is enough to overcome the fundamental rate-distance limit of QKD. Taken together, these findings suggest that our security proof method enables a versatile, fast, and possibly optimal approach towards the security validation of practical MDI-QKD systems.
\end{abstract}

\maketitle

\section{Introduction}
Quantum key distribution (QKD) is an emerging quantum technology that enables the exchange of cryptographic keys in an untrusted network \cite{bennett1984,ekert1991}. The key feature of the technology is that it is provably secure. More specifically, its security \footnote{Security in this context refers to the inaccessibility of information shared between two parties by any other individuals.} is based solely on the laws of quantum mechanics and not on how powerful the adversary could be. For this reason, secret keys generated by QKD systems can be safely used in any cryptographic applications which require long-term security assurance.

A hugely popular QKD configuration is the prepare-and-measure (P\&M) scheme. As its name suggest, P\&M-QKD involves the preparation of a quantum state that is randomly drawn from a pre-agreed set of states by the protagonist, whom we call Alice. Then, she will send the prepared state to her distant associate, whom we call Bob, to be measured in a basis that is randomly drawn from a pre-agreed set. Since Eve could not perfectly discriminate the traversing signals, any attempt to learn its identity will inevitably modify the quantum state itself. Thus, the secrecy of the secret key is guaranteed by quantum theory and such is the main appeal of QKD.

Despite the rigorous mathematical validation of QKD, present physical implementations of QKD are vulnerable to numerous side-channel attacks. In particular, it was demonstrated that the adversary could exploit the imperfections of single-photon detectors to steal some information about the secret key without introducing noise into the quantum channel \cite{fung2007,qi2007,zhao2008,xu2010,lydersen2010,gerhardt2011}. Consequently, these demonstrations necessitate the search for a QKD protocol that is robust against detection side-channel attacks: the measurement-device-independent QKD (MDI-QKD) \cite{lo2012,braunstein2012}.

\begin{figure*}
  \includegraphics[width=0.9\textwidth]{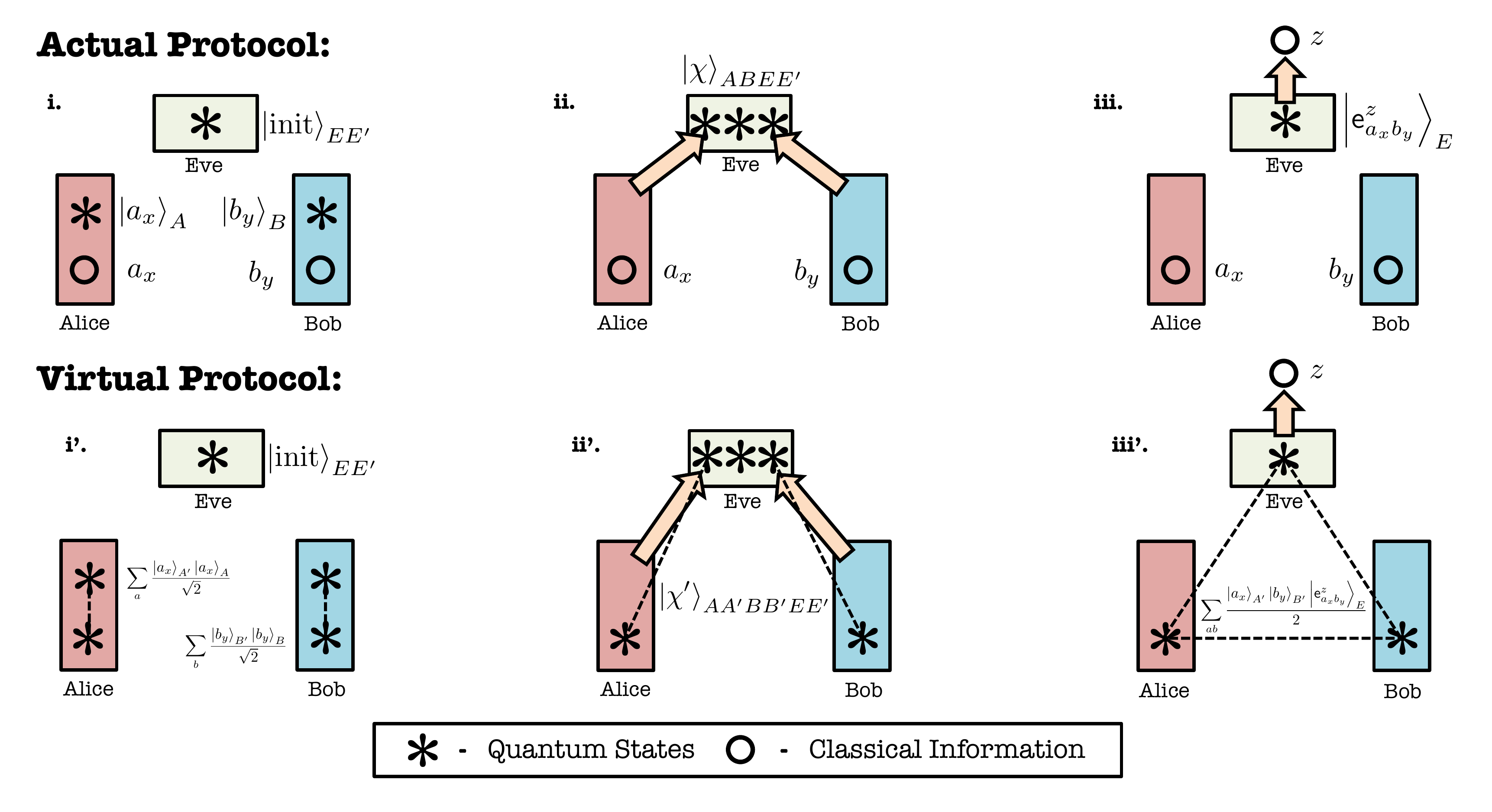}
  \caption{\textbf{Virtual protocol for arbitrary MDI-QKD protocols}. To prove the security of an MDI-QKD protocol, we consider a virtual protocol, which from the point of view of the adversary is equivalent to the actual protocol. Since the adversary could not distinguish if Alice and Bob are performing the actual or virtual protocol, the security of both protocols are equal. However, performing security proof on the virtual protocol is much easier and hence, we consider this virtual protocol. In this virtual protocol, (\emph{i'}) Alice and Bob each prepares an entangled state which can be steered into the signal states of the real protocol. Furthermore, we can choose the entangled state such that one part of the entangled state is a qubit which is kept in each party's quantum memory. (\emph{ii'}) Then, the other part of the entangled states, namely system $A$ and $B$, are sent through the quantum channel to the central node. (\emph{iii'}) When the central node performs a successful Bell state projection, Alice's qubit is now entangled with that of Bob due to entanglement swapping. As such, one can see the virtual protocol as a procedure of distributing entangled qubit pairs to Alice and Bob. Then, one can use the techniques for entanglement-based QKD to prove the security of MDI-QKD.}
  \label{fig:mdischeme}
\end{figure*}

The MDI-QKD scheme (see Fig.~\ref{fig:mdischeme}) requires both Alice and Bob to each prepare a quantum state, which, like the P\&M scheme, is randomly drawn from a pre-agreed set of states, to be sent to a potentially untrusted node connecting them. Ideally, the node will perform a Bell-state measurement jointly on the prepared states, whose outcome will determine if both Alice and Bob will keep the indices of the prepared quantum states for key distillation. This process is equivalent to a protocol where Alice and Bob each prepares a bipartite entangled state and measuring a part of it \cite{bennett1992} while sending the other to the node for entanglement swapping. Hence, it is as though Alice and Bob are making local measurements on a shared entangled state. Subsequently, Alice and Bob would sample their bit strings to evaluate their security against any potential adversary. Such security analysis is made assuming that Eve has full access and control of the node, and hence, any attack based on imperfect detectors is part of the untrusted quantum channel.

An innovation is only as good as its feasibility and the same applies to QKD protocols. With practicality in mind, MDI-QKD was specifically designed to be implementable with existing hardware, differentiating itself from other QKD protocols. Although the security of MDI-QKD relies mainly on instances when single-photon states are transmitted from Alice and Bob to the node, employing the decoy-state method \cite{lo2005decoy} gives good security performances using phase-randomized coherent states as signals. Due to its merits, efforts to progress MDI-QKD were made on both theoretical \cite{curty2014finite,tamaki2012phase,coles2016,winick2018,tamaki2014loss,ma2012statistical,xu2014protocol} and experimental \cite{rubenok2013real,liu2013experimental,da2013proof,tang2014measurement,tang2014experimental,pirandola2015high,wang2015phase,comandar2016quantum,tang2016measurement,yin2016measurement} fronts. 

Evolving from the original protocol, MDI-QKD could also be performed with coherent states without phase randomization and hence the decoy-state method. In this case, Alice and Bob encode the key information in the phases of the coherent states. The modified protocol can be proven secure using the so-called \emph{quantum coin} method (refer to Appendix \ref{section: quantum coin}) \cite{tamaki2012phase}, which was based on earlier research in P\&M-QKD with imperfect devices \cite{gottesman2004security}. However, it is not known if this approach would lead to optimal secret key rates as most security proofs provide only a lower bound on the amount of extractable secret keys without proving that such lower bound cannot be improved upon.

Here, we introduce a versatile, fast and almost-tight method to prove the security of discretely modulated MDI-QKD \footnote{Discretely modulated MDI-QKD are MDI-QKD protocols that involves finite number of different quantum states sent from Alice and Bob to the untrusted node. This is in contrast with continuous modulated states (e.g. Gaussian modulated states), which in practice are always approximated by a discretely modulated state.}. This technique is based on the observation that the Gram matrix of Eve's quantum side information \footnote{Since Eve's quantum side information is described by quantum states, a matrix with elements as the inner-products of these states, known as a Gram matrix, must be positive semi-definite.} translates to the characterization of Eve's attack in a compact way (refer to section~\ref{section:method} for more details). This allows us to compute the phase error directly, which makes our method almost-tight, as compared to the pessimistic estimation obtained using existing technique \cite{tamaki2012phase}. We apply this characterization technique to three different types of MDI-QKD protocols: (1) the original polarization (or time-phase) encoding decoy-state MDI-QKD \cite{lo2012}, (2) phase-encoding MDI-QKD using non-phase-randomized coherent states \cite{tamaki2012phase}, and (3) a variant of twin-field QKD \cite{lucamarini2018} called phase-matching MDI-QKD \cite{lin2018}. The proposed proof technique demonstrates significant advantages over the known bounds on secret key rates for all of the considered MDI-QKD protocols.

\section{Method}
\label{section:method}
In this section, we outline a general security analysis procedure that works for any MDI-QKD protocol, provided the encoded states are pure and their inner-products are known. In practice, one can characterize the encoded states by performing state tomography and thus their inner-products can be easily obtained. For simplicity, we only consider security against collective attacks in the asymptotic limit in this paper. To extend our result to security against coherent attacks in the finite-key regime, we can apply the post-selection technique \cite{christandl2009postselection} or the entropy accumulation theorem \cite{dupuis2016entropy,arnon2018practical}.

Suppose the states that Alice prepares are $\{\ket{\As(a_x)}_A\}_{a_x}$ and those prepared by Bob are $\{\ket{\Bs(b_y)}_B\}_{b_y}$ where $a$ and $b$ are the bit values while $x$ and $y$ are the basis choices of Alice and Bob respectively. Furthermore, Alice's and Bob's quantum encoding, denoted by $\ket{\As(\cdot)}$ and $\ket{\Bs(\cdot)}$ respectively, can be different in general. We denote the inner-product of the code states by $\Lambda_{a_x b_y,a'_{x'} b'_{y'}} = \braket{\As(a_x)}{\As(a'_{x'})}_A \braket{\Bs(b_y)}{\Bs(b'_{y'})}_B$. Note that $\Lambda_{a_x b_y,a'_{x'} b'_{y'}}$ could be complex.

Since the joint input states are pure, we can describe the quantum channel and the untrusted measurement by the central node as a quantum-to-classical map. Such maps can be described by an isometry $\U$ acting on the input states giving
\begin{equation}
    \ket{\As(a_x)}_A \ket{\Bs(b_y)}_B \xrightarrow{\U} \sum_{z} \ket{\Es_{a_x b_y}^z}_E \ket{z}_{E'}
\end{equation}
where $z$ is the classical announcement made by the central node. For concreteness, we consider the case where $z \in \{\pass, \fail \}$, which indicates whether the Bell state measurement is successful or not, however, our method allows for other announcements by the central node, such as the actual result of the Bell state measurement. The vectors $\{\ket{\Es_{a_x b_y}^z}_E\}_{a_x,b_y,z}$ are sub-normalized states corresponding to the quantum side information that Eve holds. Now, one can construct a Gram matrix $G$ for the adversary's quantum side information and it is clear that such matrix is positive semi-definite, i.e., $G \succeq 0$.

Moreover, since the channel is described by an isometric transformation, the inner-product information is preserved at the output states as well. More specifically, we have the following constraint on the Gram matrix $G$:
\begin{align}
    \Lambda_{a_x b_y,a'_{x'} b'_{y'}}
    &= \braket{\As(a_x)}{\As(a'_{x'})}_A \braket{\Bs(b_y)}{\Bs(b'_{y'})}_B \nonumber\\
    &= \sum_z \braket{\Es_{a_x b_y}^z}{\Es_{a'_{x'} b'_{y'}}^z}_E
\end{align}
Furthermore, other observable constraints such as the bit-error rate and the probability of a successful Bell state measurement can be expressed in terms of linear constraints on the Gram matrix. For example, the
probability of successful Bell state measurement when Alice and Bob both choose the basis $\gamma$, denoted by $P_\pass^{\gamma}$, is given by
\begin{align}
    P_\pass^{\gamma}
    &= \sum_{a,b} \frac{P(a_\gamma,b_\gamma)}{f_\gamma} P(\pass|a_\gamma,b_\gamma)\nonumber\\
    &= \sum_{a,b} \frac{P(a_\gamma,b_\gamma)}{f_\gamma} 
    \braket{\Es_{a_\gamma b_\gamma}^\pass}{\Es_{a_\gamma b_\gamma}^\pass}_E
\end{align}
where $f_\gamma$ is the probability of Alice and Bob both choosing the $\gamma$ basis and $P(a_\gamma, b_\gamma)$ is the joint probability that Alice chooses $a_\gamma$ and Bob chooses $b_\gamma$. Similarly, the bit-error rate in the $\gamma$ basis, denoted by $\ebit^\gamma$, can be expressed as
\begin{align}
    \ebit^\gamma &= P(a \neq b | \pass, x=y=\gamma) \nonumber\\
    &= \sum_{a \neq b} \frac{P(a_\gamma,b_\gamma)}{P_\pass^{\gamma} f_\gamma}
    P(\pass | a_\gamma, b_\gamma) \nonumber\\
    &= \sum_{a \neq b} \frac{P(a_\gamma,b_\gamma)}{P_\pass^{\gamma} f_\gamma}
    \braket{\Es_{a_\gamma b_\gamma}^\pass}{\Es_{a_\gamma b_\gamma}^\pass}_E
\end{align}
where the first equality is from the definition of bit-error rate and the second equality can be easily obtained by applying Bayes rule.

Now, to prove the security of the protocol, we consider a virtual protocol illustrated in Fig. \ref{fig:mdischeme}. Since the adversary could not distinguish if Alice and Bob are performing the actual or virtual protocol, the security of both protocols are equal. However, performing security proof on the virtual protocol is much easier and hence, we consider this virtual protocol. For concreteness, we suppose that Alice and Bob extract secret key from the $\K$ basis and they use the $\T$ basis (which is mutually-unbiased to $\K$) to estimate Eve's information. Note that these bases are defined in the virtual protocol and not in the actual protocol. In this virtual protocol, to generate secret key, Alice and Bob first prepare the entangled states
\begin{align}
    \ket{\Psi_\K}_{A'A} = \frac{1}{\sqrt{2}}
    \left[\sum_a \ket{a_\K}_{A'}\ket{\As(a_\K)}_A\right] \nonumber\\
    \ket{\Psi'_\K}_{B'B} = \frac{1}{\sqrt{2}}
    \left[\sum_b \ket{b_\K}_{B'}\ket{\Bs(b_\K)}_B\right]
\end{align}
where the systems $A'$ and $B'$ are qubits. From the point of view of Eve, this is equivalent to the actual protocol since after tracing out systems $A'$ and $B'$, the marginal states are the same as if Alice and Bob prepare the states $\ket{\As(a_\K)}_A$ and $\ket{\Bs(b_\K)}_B$ randomly based on the random bit $a$ and $b$ given by their random number generator. Furthermore, Alice and Bob can obtain the random bit $a$ and $b$, simply by measuring systems $A'$ and $B'$ in the $\K$ basis.

After the central node successfully performs a Bell state projection, the qubit-qubit system $A'B'$ are now entangled due to the entanglement swapping \footnote{Entanglement swapping is a technique used for distributing entangled states between two distant parties via a node between them. By establishing entanglement between each party and the node, the node could entangle the quantum states held by the two parties by performing a Bell-state measurement and announcing the measurement outcome. }\cite{bennett1993teleporting,zukowski1993event} operation. Depending on the outcome of the Bell state measurement $z$, Bob will need to perform a local unitary $\mathbb{I}_{A'} \otimes U_{B'}(z)$ to rotate the state into the target state. Then, to estimate the side information that Eve has, Alice and Bob can measure the systems $A'$ and $B'$ respectively in the $\T$ basis. This will give Alice and Bob the so-called phase-error rate $\eph$ in their virtual qubit-qubit system. Shor and Preskill \cite{shor2000} proved that, for this type of QKD protocol, the lower bound on the asymptotic secret key rate is given by
\begin{equation}
    R \geq P_\pass^\K \Big[ 1 - h_2(\eph) - h_2(\ebit) \Big]
\end{equation}
where $h_2(\cdot)$ is the binary entropy function.

At this point, a few remarks are in order. Firstly, the measurement in the $\T$ basis when Alice and Bob prepare the entangled states $\ket{\Psi_\K}_{A'A}$ and $\ket{\Psi'_\K}_{B'B}$ are counter-factual measurements that are not performed in the actual protocol. The actual test states that Alice and Bob send are $\{\ket{\As(a_{\T})}_A\}_a$ and $\{\ket{\Bs(b_{\T})}_B\}_b$ whose virtual analogue is measuring the qubit systems of some other entangled states $\ket{\Psi_{\T}}_{A'A}$ and $\ket{\Psi'_{\T}}_{B'B}$ in the $\T$ basis. As such, the phase-error rate $\eph$ is not accessible in the actual protocol. However, Alice and Bob can employ the knowledge of the test states (i.e., the inner-product information), the observed error rate, and the success probability to constrain the virtual protocol phase-error rate, $\eph$, as we will show later.

Secondly, like the bit-error rate $\ebit$, the phase-error rate $\eph$ is also a linear function of the Gram matrix $G$. The explicit expression of $\eph$, in general depends on the choice of entangled state that Alice and Bob prepares, the virtual measurements that steer the states, as well as the target state of the qubit pairs shared between Alice and Bob. As an example, the procedure to obtain the explicit expression for $\eph$ in the phase-encoding MDI-QKD protocol is outlined in Appendix \ref{section: Appendix phase error}.

With all these in mind, here is the crucial observation. The Gram matrix $G$ depends on the isometry $\U$ performed by Eve. Secondly, the constraints on the Gram matrix $G$, imposed by the inner-product information and the observed statistics, are linear. Moreover, one can obtain a lower bound on the secret key rate $R$ by obtaining an upper bound on the phase-error rate $\eph$, which is a linear objective function, allowing us to use semi-definite programming (SDP) techniques to completely characterize Eve's strategies. This will be explained in more details in section \ref{section: sdp}.

Lastly, to show that our method gives a tighter lower bound on the key rate as compared to the quantum coin approach, we will analyze MDI-QKD protocols based on coherent states and Trojan horse attacks. These are the scenarios in which the quantum coin approach was used in the literature. The outcomes of the simulation are presented in section \ref{section: examples}.

\section{Characterizing Eve's strategy using semi-definite programming} \label{section: sdp}

SDP is a class of convex optimization problem with a linear objective function over a cone of positive semi-definite matrices \cite{boyd2004convex}. One can think of such optimization problem as linear programming with linear matrix inequalities as constraints. Just like any other optimization problems, we can define the primal problem and the dual problem of an SDP. If the primal problem is a minimization problem, then its dual would be a maximization problem and \textit{vice versa}. Furthermore, if the primal problem is a maximization problem, then the weak duality theorem states that the optimal primal value $p^*$ and the optimal dual value $d^*$ satisfy the following relationship
\begin{equation}
    p^* \leq d^*
\end{equation}
and the opposite is true if the primal problem is a minimization problem. When $p^* = d^*$, we say that strong duality holds and it implies that $p^* = d^*$ is the optimal solution to the SDP.

Previously, there have been other numerical approaches to security analysis of QKD \cite{wang2019characterising, coles2016, winick2018}. One of the appeals of such numerical approaches is that they are highly versatile and can be used to analyze a wide range of QKD protocols without relying on the specific structure of the underlying design. Here we propose an SDP-based technique of analyzing the security of MDI-QKD that is versatile, simple and almost-tight. The notion of almost-tight refers to the obtained upper and lower bounds on the phase error, $\eph$, coincide up to some numerical precision, which is inherent to any numerical methods.

Despite also approaching the security analysis numerically, we emphasize that our method is conceptually distinct from previous methods presented in Refs. \cite{coles2016,winick2018,wang2019characterising}. In Refs. \cite{coles2016,winick2018}, one would lower bound the secret key rate by bounding the von Neumann entropies in the Devetak-Winter bound \cite{devetak2005distillation} directly. In Ref. \cite{coles2016}, this was done by applying the Golden-Thompson inequality to make the problem more manageable, whereas in Ref. \cite{winick2018}, one would have to find an almost-optimal attack by the eavesdropper and then linearize the objective function and solve the optimization problem using SDP. For Ref. \cite{wang2019characterising}, one bound the phase-error rate instead of the von Neumann entropy. This was done by introducing a hierarchy of semi-definite relaxations that bounds the set of quantum correlations from outside and express the phase-error rate as a linear objective function. Once a linear objective function is obtained, one can use any SDP solver to obtain the secret key rate.

Like Ref. \cite{wang2019characterising}, our method computes the secret key rate by bounding the phase-error rate $\eph$. Our method exploits the fact that the Gram matrix $G$ is both positive semi-definite and linearly constrained. As such, obtaining an upper bound to the phase-error rate is equivalent to solving the dual problem of an SDP. In other words, the solution of the SDP characterizes the optimal attack that Eve can perform on the quantum channel and with the untrusted central node. Thus, the security analysis of any MDI-QKD protocol is equivalent to solving the following SDP
\begin{eqnarray} \label{eq: our SDP}
	&\mathtt{maximize}&~ \eph \nonumber\\
	&\mathtt{s.t}&~ G \succeq 0 \nonumber\\
	&&~ \eph \leq 1/2 \nonumber\\
	&&~ P_\pass^\gamma = \sum_{a,b} \frac{P(a_\gamma,b_\gamma)}{f_\gamma} 
    \braket{\Es_{a_\gamma b_\gamma}^\pass}{\Es_{a_\gamma b_\gamma}^\pass}_E \nonumber\\
	&&~ \ebit^\gamma = \sum_{a \neq b} \frac{P(a_\gamma,b_\gamma)}{P_\pass^{\gamma} f_\gamma}
    \braket{\Es_{a_\gamma b_\gamma}^\pass}{\Es_{a_\gamma b_\gamma}^\pass}_E \nonumber\\
	&&~ \Lambda_{a_x b_y,a'_{x'} b'_{y'}} = \sum_z \braket{\Es_{a_x b_y}^z}{\Es_{a'_{x'} b'_{y'}}^z}_E
\end{eqnarray}
where $\gamma$ is the basis choice of Alice and Bob and $\eph$ is a linear function of the Gram matrix $G$, which is protocol dependent.  In order to obtain a reliable upper bound on $\eph$, one has to optimize the dual problem of optimization~\eqref{eq: our SDP}.

Furthermore, due to the weak duality theorem, any feasible solution of the dual problem gives a reliable upper bound on the phase-error rate. Moreover, when the strong duality condition holds, our upper bound on the phase-error rate is tight. Remarkably, in practice, the duality gap (the difference between the primal and dual value) is usually very small. Hence, our upper bound on the phase-error rate is almost tight, even when the strong duality condition is not met.

Drawing the connection to the isometry $\U$ performed by the eavesdropper, since the Gram matrix characterizes $\U$, solving the SDP \eqref{eq: our SDP} is equivalent to finding the isometry that gives Eve the maximum knowledge about the bits that Alice and Bob possess, considering the statistics that Alice and Bob observe. Lastly, we remark that SDPs can be efficiently solved with standard solvers that are readily accessible. Therefore, our method can be implemented easily even with personal computers (which is the case for the simulations presented in this work).
\begin{figure*}[t!] 
	\includegraphics[width=\textwidth,trim = {0 6.5cm 0 0}]{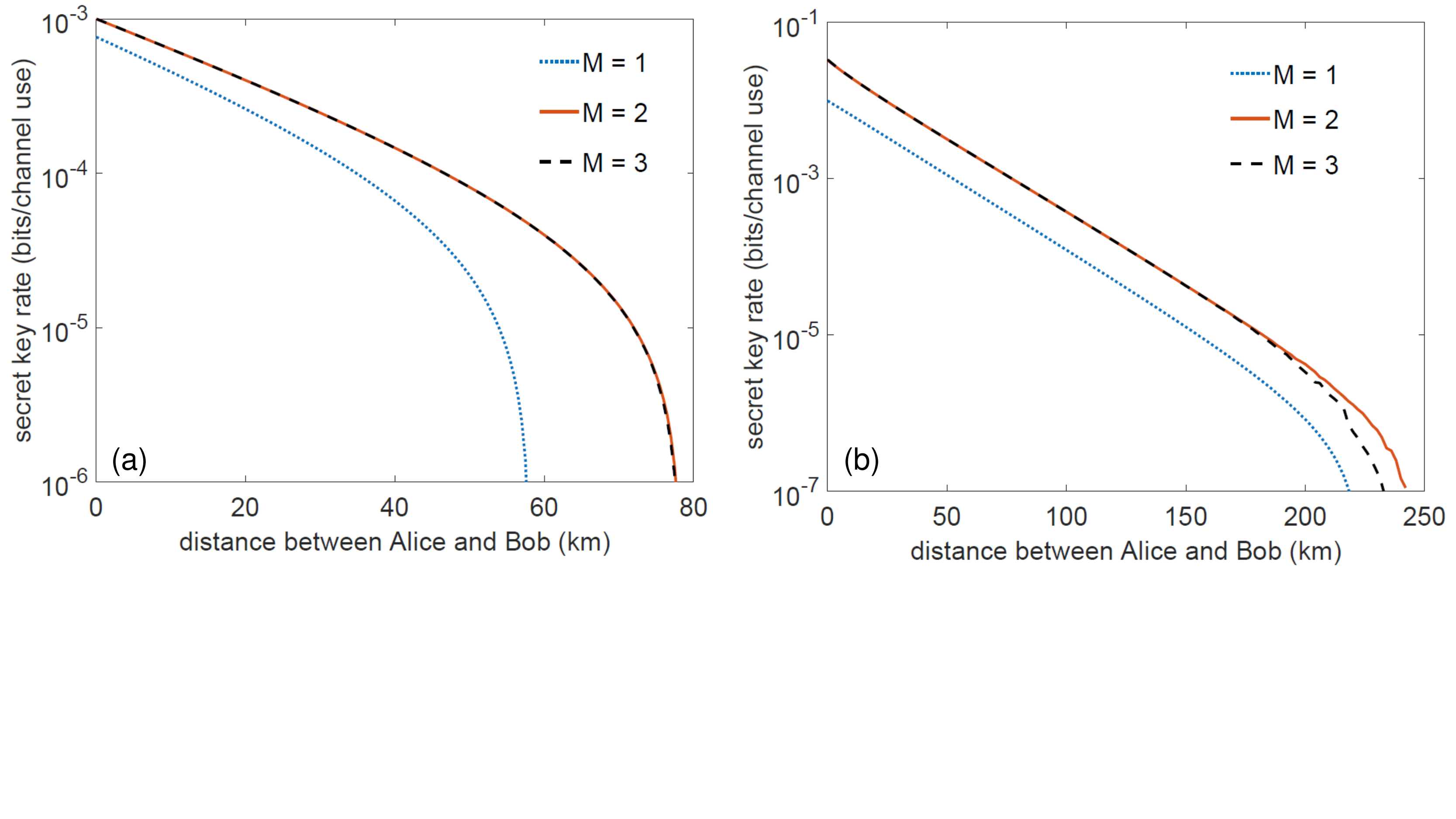}
	\caption{\textbf{Key rate simulation with (a) APD and (b) SNSPD for generalized phase-encoding coherent state MDI-QKD}. We use our method to compute the lower bound on the secret key rate for the one-basis (blue-dotted curves), two-basis (red-solid curves) and three-basis protocols (black-dashed curves). As one can see, the curves corresponding to the three-basis protocol almost coincide with the curves corresponding to the two-basis protocol.}
	\label{fig: coherent state MDI key rate}
\end{figure*}

At this point, we highlight a few differences between our method and other numerical methods that have been proposed previously. Firstly, our method does not rely on any semi-definite relaxation unlike Ref. \cite{wang2019characterising}. It is also important to note that Ref. \cite{wang2019characterising} did not formulate any framework to apply their technique to MDI-QKD. Moreover, compared to the methods of Refs. \cite{coles2016, winick2018}, which involves nonlinear optimization, our method has significant advantage in terms of computational time. This gives a practical appeal when one use the numerical toolbox for optimization of protocol parameters. Additionally, the application of Golden-Thompson inequality in Ref. \cite{coles2016} may affect the convexity of the optimization problem when the inequality is strict. This can affect both the efficiency and optimality of the method as the algorithm might yield a locally optimal result instead of the global one.

\section{Examples} \label{section: examples}
In this section, we apply our technique to coherent state based protocols, namely the phase-encoding coherent state MDI-QKD \cite{tamaki2012phase} and a variant of the recently proposed twin-field QKD (TF-QKD) protocol \cite{lucamarini2018}, as well as the decoy-state protocols \cite{lo2012} with Trojan horse attacks. We also analyze these protocols with the quantum coin method and show that our technique achieves significantly higher secret key rate. For simplicity, we assume that the central node is equidistant to Alice and Bob. Consequently, we can also assume that the set of states that Alice prepares is the same to that of Bob. Although these assumptions are not necessary for our method to work, they will simplify the calculations greatly.

For the simulations of phase-encoding coherent state MDI-QKD as well as the decoy-state MDI-QKD, the parameters that we use are listed in Table \ref{table: parameters}. Parameter 2 will also be used for the simulation of TF-QKD.
\begin{table}[ht]
\centering
\caption{\textbf{List of parameters for numerical simulations.} The different parameters corresponds to different type of detectors commonly used in QKD experiments. Parameter 1 corresponds to the case when \textit{avalanche photodiode} (APD) detectors are used. Typically, APD detectors have relatively high dark count rate $p_{dc}$ and low detector efficiency $\eta_{\text{det}}$. On the other hand, Parameter 2 corresponds to the case where  \textit{superconducting nanowire single-photon-detectors} (SNSPD) are used. SNSPDs typically have lower dark count rate and higher detector efficiency. For all the simulations presented in this paper, we assume that the single mode fiber has loss coefficient $\xi = 0.2 $dB/km and the overall misalignment error in the channel $e_\text{ali} = 1.5\%$. Furthermore, we assume that error corrections can be done at Shannon's limit. Then, for each given distance, we optimize the intensity of the signals to get the highest lower bound on the secret key rate.}
 \begin{tabular}{|r || c c c c|} 
 \hline
  & $p_{dc}$ & $\eta_{\text{det}}$ & $\xi$ (dB/km) & $e_{\text{ali}}$ \\ [0.5ex] 
 \hline
 Parameter 1  & $6.02 \times 10^{-6}$ & 14.5\% & 0.20 & 1.5\% \\
 Parameter 2  & $5 \times 10^{-8}$ & 85\% & 0.20 & 1.5\% \\[1ex]
 \hline
 \end{tabular}
\label{table: parameters}
\end{table}

\subsection{Phase-encoding coherent state MDI-QKD} \label{subsection: phase-encoding}

We first analyze the phase-encoding coherent state MDI-QKD protocols. In this protocol, Alice and Bob independently send non-phase-randomized weak coherent pulses of fixed intensities which are phase-modulated. The central node would then interfere the signals that they send using a beam-splitter and then performs single photon detection on both outputs of the beam-splitter. When exactly one of the two detectors clicks, Alice and Bob will be able to learn whether their signals are correlated or anti-correlated. With a simpler architecture and fewer components and their related flaws, this class of MDI-QKD protocols can, in principle, provide economic and high-speed QKD systems. However, they have not been implemented experimentally due to other technical challenges such as phase-locking two independent lasers at a distance. Furthermore, based on the quantum coin approach, the secret key rate achieved by these protocols is not favorable. However, the interest in coherent state MDI-QKD protocols has been revived recently due to the introduction of twin-field-QKD (TF-QKD) protocols, also known as phase-matching MDI-QKD \cite{lucamarini2018, ma2018, tamaki2018information, cui2019twin, curty2018simple, lin2018}. Remarkably, TF-QKD enables the distribution of secret key with rate that overcomes the so-called repeaterless bound \cite{takeoka2014, pirandola2017} due to single-photon interference. As such, we foresee that there will be endeavours in implementing phase-locking techniques in QKD systems in the near future. Furthermore, as we shall see, we can prove that the secret key rate of coherent state protocols can be significantly improved using our method. Interestingly, as we demonstrate later, the secret key rate phase-encoding coherent state protocols can even be higher than that of standard decoy-state MDI-QKD protocols at short distances. Therefore, it will be interesting to study this class of protocols.

In Ref.~\cite{tamaki2012phase}, Alice and Bob prepare the states $\{ \ket{\pm \sqrt{\mu}}, \ket{\pm i \sqrt{\mu}}\}$ where $\mu$ is the mean photon number of the coherent states. They use the basis $\{\ket{+\sqrt{\mu}}, \ket{-\sqrt{\mu}}\}$ to generate key and the basis $\{\ket{+i \sqrt{\mu}}, \ket{-i \sqrt{\mu}}\}$ for parameter estimation. On the other hand, the untrusted central node will announce $z \in \{ \Psi^+, \Psi^-, \varnothing \}$ depending on whether the outcome of the Bell state measurement is $\Psi^+$, $\Psi^-$ or unsuccessful respectively. Furthermore, to improve the efficiency of the protocol, whenever the central node announces $z = \Psi^-$, Bob will flip the value of his bit.  

We note that in Bell state measurements of coherent states, the outcome $\Phi^+$ cannot be distinguished from the outcome $\Psi^+$ and similarly, the outcome $\Phi^-$ cannot be distinguished from the outcome $\Psi^-$ as mentioned in Ref. \cite{tamaki2012phase}. Consequently, there will be intrinsic phase-error rate even if the bit-error rate is zero. However, the probability of obtaining $\Psi^+$ and $\Psi^-$ is higher than the probability of obtaining $\Phi^+$ and $\Phi^-$ respectively when the intensity of the coherent states is small (which is typically the case for QKD). Hence, we choose to label the conclusive outcomes as $\Psi^+$ and $\Psi^-$.

Here, we generalize the protocol to allow Alice and Bob to choose from $M$ possible bases. Thus, the protocol proposed in Ref. \cite{tamaki2012phase} corresponds to the case where $M = 2$. The case where $M = 1$ is known in the literature as the MDI-B92 protocol and was studied in Ref. \cite{ferenczi2013security}. In this section, we will also study the case where $M=3$.

For the generalized protocol, Alice and Bob independently prepare one of the states of the form
\begin{align}
    \ket{\As_{a,\theta_x}} =  \ket{(-1)^{a} e^{i \theta_x} \sqrt{\mu}} \nonumber\\
    \ket{\Bs_{b,\theta_y}} = \ket{(-1)^{b} e^{i \theta_y} \sqrt{\mu}}
\end{align}
where $a,b \in \{0,1\}$ is her (his) bit value, $\theta_x = x\pi/M$ and $\theta_y = y\pi/M$ with $x,y \in \{0,1, ... ,M-1\}$ as her (his) choice of basis.

In the following simulations, for each given distance, we choose the value of the mean photon number $\mu$ that maximize the lower bound on the secret key rate. The result of the simulation is shown in Fig. \ref{fig: coherent state MDI key rate}

Surprisingly, the three-basis protocol does not give noticeable improvement from the two-basis protocol in terms of the lower bound on the secret key rate. The intuition behind this is that one can estimate the statistics of the 4 test states in the three-basis protocol with the 2 test states in the two-basis protocol when we require all the intensities of the code states to be the same. Interestingly, when we remove that constraint, increasing the number of test states can give significant improvement in the secret key rate, as we shall see when we discuss about TF-QKD later.

\begin{figure*}[t!] 
	\includegraphics[width=\textwidth, trim = {0 6.5cm 0 0}]{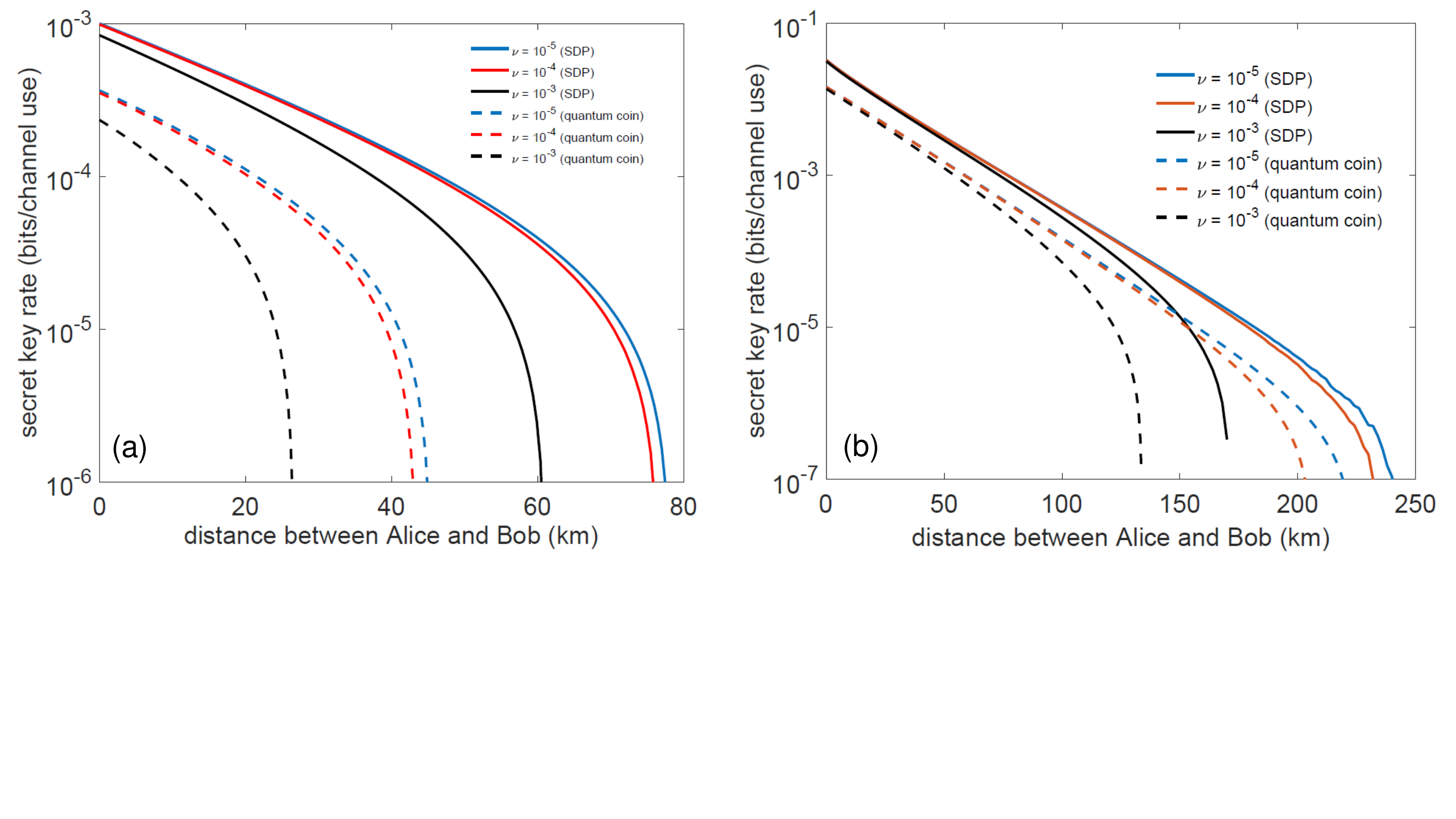} 
	\caption{\textbf{Key rate simulation with (a) APD and (b) SNSPD for phase-encoding coherent state MDI-QKD under Trojan horse attack}. We compare the secret key rate obtained using our method (shown by solid curves) with the one obtained using the quantum coin technique (shown by dashed curves). For the simulation, we assume different intensities of the leakage light $\nu$ (From top to bottom: $\nu = 10^{-5}$ (blue), $\nu = 10^{-4}$ (red) and $\nu = 10^{-3}$ (black)). Comparing the two methods, we can see that for a given intensity of leakage, the secret key rate computed using our method is consistently higher than the secret key rate obtained by the quantum coin approach. Furthermore, one can see that the key rate obtained by our method is more robust against Trojan horse attacks.}
	\label{fig: coherent state MDI with THA keyrate}
\end{figure*}

To compare our method with the quantum coin method, we consider the two-basis protocol. Additionally, we will also evaluate the performance of the two techniques when Eve performs Trojan horse attack on the sources of Alice and Bob. The Trojan horse attack is executed by shining a bright light to Alice's or Bob's source. Some of the light would be reflected and leak some information to the adversary as it also passes through the modulator. For the model of the Trojan horse attack, we extend the model proposed by Ref. \cite{lucamarini2015practical}. The authors of Ref. \cite{lucamarini2015practical} consider Trojan horse attack on prepare-and-measure QKD but it is rather straightforward to extend the model to MDI-QKD where Eve attacks both Alice's and Bob's sources. We assume that in executing the Trojan horse attack, Eve injects coherent pulses into Alice's and Bob's sources and they modulate the reflected coherent light with intensity $\nu$. In the presence of Trojan horse attack, the output of Alice's source can be written as
\begin{align}
    \ket{\As_0}_A &=
    \ket{+\sqrt{\mu} }_{\bar{A}} \otimes \ket{+\sqrt{\nu}}_{T_A} \nonumber\\
    \ket{\As_1}_A &=
    \ket{-\sqrt{\mu} }_{\bar{A}} \otimes \ket{-\sqrt{\nu}}_{T_A} \nonumber\\
    \ket{\As_2}_A &=
    \ket{+i\sqrt{\mu} }_{\bar{A}} \otimes \ket{+i\sqrt{\nu}}_{T_A} \nonumber\\
    \ket{\As_3}_A &=
    \ket{-i\sqrt{\mu} }_{\bar{A}} \otimes \ket{-i\sqrt{\nu}}_{T_A}
\end{align}
where $\nu$ is the intensity of the reflected light. Here, for simplicity of notation, we replace the pair $(a,\theta_x)$ with a single index. We assume that the Trojan horse system $T_A$ is only accessible to Eve and the probability of a successful Bell state measurement $P_\pass$ is not affected by the Trojan horse attack. The output of Bob's source is modeled similarly. The results of the simulations are shown in Fig. \ref{fig: coherent state MDI with THA keyrate}.

As we can see from the figure, our method gives a tighter bound on the secret key rate than the quantum coin method. This is because the inequality that was used in the quantum coin approach is, in general, not tight. On the contrary, our SDP gives an almost tight approximation of the phase-error rate, owing to the small duality gap. Furthermore, our method is also more robust against the Trojan horse attack. This is because for a given distance, our method predicts higher optimal intensity than the one predicted by the quantum coin method. As a result, the ratio of the intensity of the signal states $\mu$ to the intensity of the leaked light $\nu$ is higher when computed with our method. Remarkably, even in the absence of Trojan horse attack (when $\nu = 0$), our method gives a significantly higher lower bound on the secret key rate.

\subsection{Decoy-state MDI-QKD with Trojan horse attacks}
We now consider the decoy-state MDI-QKD protocol \cite{lo2012}. For concreteness, we consider the time-phase encoding but our results can be transferred to other encoding scheme such as the polarization encoding. In this protocol, Alice and Bob randomly modulate the intensities of their lasers. For each channel use, they will phase-randomize their laser such that the output of their sources can be described by Poissonian distributions of photon number states with different means determined by their respective intensity settings. In the post-processing step, Alice and Bob then can estimate the gain and error rate, given their intensity settings. In turn, they can estimate the single-photon yield and error rate.

\begin{figure*}[t!] 
	\includegraphics[width=\textwidth, trim = {0 6.5cm 0 0}]{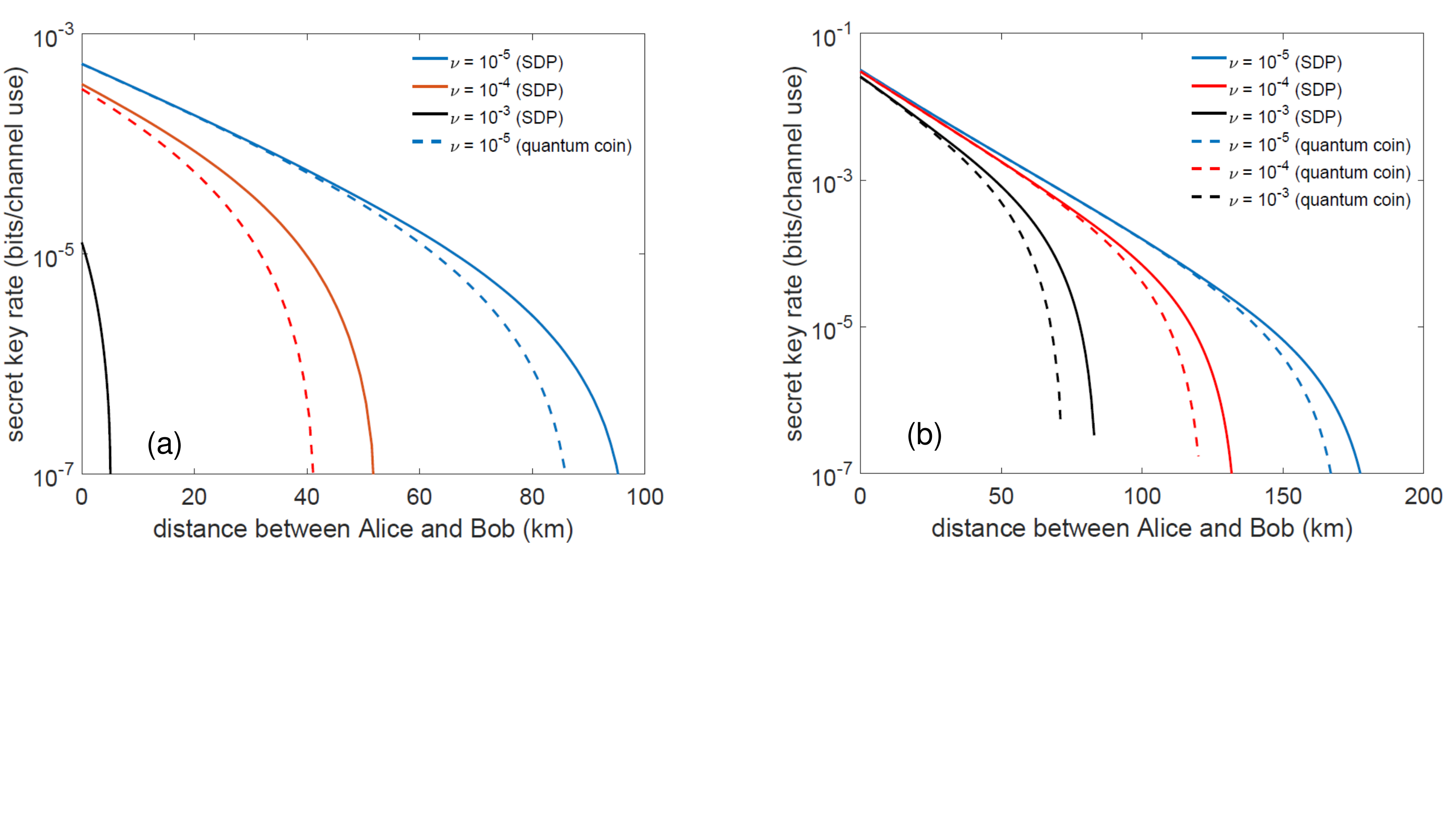} 
	\caption{\textbf{Key rate simulation with (a) APD and (b) SNSPD for time-phase-encoding decoy-state MDI-QKD under Trojan horse attack}. We compare the secret key rate obtained using our method (shown by solid curves) with the one obtained using the quantum coin technique (shown by dashed curves). For the simulation, we assume different intensities $\nu$ for the Trojan horse light. (From top to bottom: $\nu = 10^{-5}$ (blue), $\nu = 10^{-4}$ (red) and $\nu = 10^{-3}$ (black)). Comparing the two methods, we can see that the secret key rate computed using our method is consistently higher than the secret key rate obtained by the quantum coin approach. We note that for the simulation with APD, when $\nu = 10^{-3}$, the quantum coin method fails to obtain any positive key rate.}
	\label{fig: decoy-state MDI keyrate}
\end{figure*}

Combined with the measurement-device-independent architecture which protects us against any detector side-channel attacks, the decoy-state method \cite{lo2005decoy} prevents a certain side-channel attack on the sources of Alice and Bob, namely the photon-number-splitting (PNS) attack \cite{huttner1995quantum,brassard2000limitations,lutkenhaus2002}. The decoy state method uses multiple intensities to approximate the single-photon yield and error rate, hence revealing PNS attacks when Eve actually performs them. Nevertheless, decoy-state MDI-QKD is still vulnerable to other attacks on the sources. One possible attack that Eve can perform with current technology is the Trojan horse attack. As such, it would be interesting to study how decoy-state MDI-QKD protocols fare when subjected to practical Trojan horse attacks.

For simplicity, we follow Ref. \cite{lucamarini2015practical}'s assumption that the Trojan horse attack does not affect the decoy-state implementation but only leaks the information about which of the signal states are being prepared by Alice and Bob. However, we note that our method can also be applied when the Trojan horse leaks the information of the decoy setting as discussed in Ref. \cite{tamaki2016decoy}. Such technical extension is out of the scope of this paper and will be left for future work.

To apply our technique to decoy-state MDI-QKD, we first estimate the single-photon yield $Y^\gamma_{11}$ as well as the single-photon quantum bit-error rate (QBER) $e^\gamma_{11}$ where $\gamma \in \{Z,X\}$ denotes the basis choice of Alice and Bob. Then, we can apply our method to estimate the single-photon phase-error rate $\eph^{11}$ and hence, the secret key rate. In the absence of Trojan horse attack, the single-photon phase-error rate $\eph^{11}$ is simply the single-photon QBER in the $X$ basis, $e^X_{11}$. Nevertheless, when Eve performs Trojan horse attack, the single-photon phase-error rate is no longer measured by Alice and Bob and can only be estimated using techniques like the quantum coin method or our proposed method. Here, we show that our method gives a tighter estimation of the single-photon phase-error rate than the quantum coin method.

Following the model of Ref. \cite{lucamarini2015practical}, Alice's single-photon signals when affected by the Trojan horse attack can be written as
\begin{align}
    \ket{\As_0}_A &=
    \ket{0}_{\bar{A}} \otimes \ket{\sqrt{\nu}}_{T_A} \ket{\vac}_{T'_A} \nonumber\\
    \ket{\As_1}_A &=
    \ket{1}_{\bar{A}} \otimes \ket{\vac}_{T_A} \ket{\sqrt{\nu}}_{T'_A} \nonumber\\
    \ket{\As_2}_A &=
    \ket{+}_{\bar{A}} \otimes \ket{\sqrt{\frac{\nu}{2}}}_{T_A} \ket{\sqrt{\frac{\nu}{2}}}_{T'_A} \nonumber\\
    \ket{\As_3}_A &=
    \ket{-}_{\bar{A}} \otimes \ket{\sqrt{\frac{\nu}{2}}}_{T_A} \ket{-\sqrt{\frac{\nu}{2}}}_{T'_A}
\end{align}
where $\{\ket{0},\ket{1}\}$ is the canonical basis and $\ket{+} = \frac{\ket{0}+\ket{1}}{\sqrt{2}}$, $\ket{-} = \frac{\ket{0}-\ket{1}}{\sqrt{2}}$ and $\ket{\vac}$ is the vacuum state. Bob's single-photon signals can be written down in similar fashion.

For the simulation, suppose that two decoy-states are used alongside the signal state. At the end of the protocol, Alice and Bob will have an estimate of the total gain $Q_{\mu_A \mu_B}^\gamma$ and the total QBER $E^\gamma_{\mu_A \mu_B}$ where $\mu_A, \mu_B \in \{\mu, \zeta, \omega\}$ are the intensity setting of Alice and Bob respectively. To simulate the statistics observed in the experiment, we follow Ref. \cite{ma2012alternative}'s model for the total gain and total QBER. Then, we use the Gaussian elimination method to estimate the single-photon yield and QBER. Assuming that $\mu \geq \zeta \geq \omega$, the lower bound on the single-photon yield $Y^{\gamma,L}_{11}$ and the upper bound on the single-photon QBER $e^{\gamma,U}_{11}$ is given in Ref. \cite{xu2014protocol}. Other methods such as linear programming will yield similar results. 

\begin{figure*}[t!] 
	\includegraphics[width=\textwidth, trim = {0 6.5cm 0 0}]{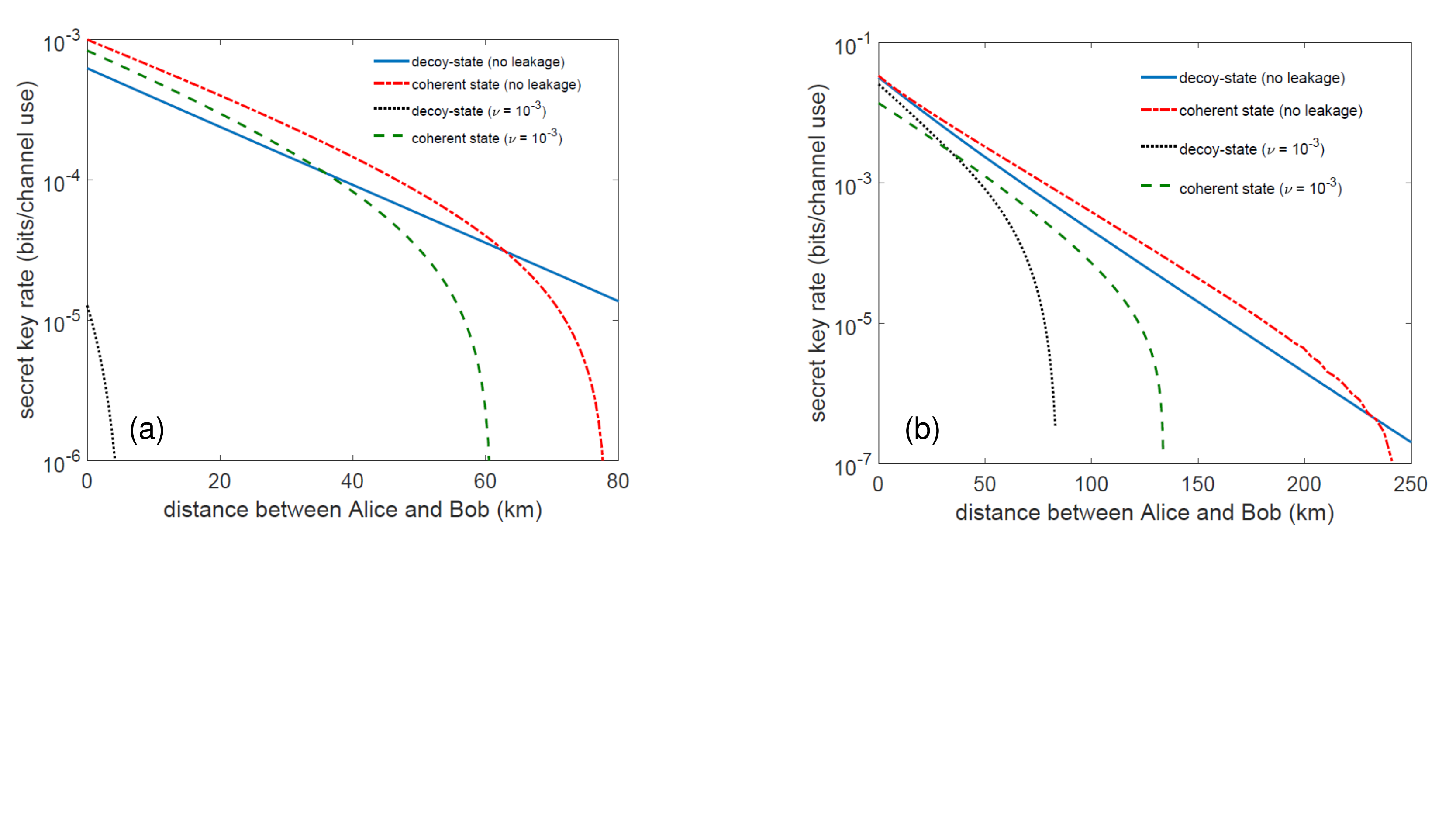} 
	\caption{\textbf{Comparison of the secret key that can be distilled  per channel use in decoy-state versus coherent state MDI-QKD}. We compare the secret key rate of time-phase encoding decoy-state and phase-encoding coherent state MDI-QKD protocols both with and without Trojan horse attack with the simulation parameters that correspond to (a) APD and (b) SNSPD. When there is no Trojan horse leakages, we can see that the coherent state protocol (shown by red-dash-dotted curves) has higher secret key rate than the decoy-state protocol (shown by blue-solid curves) at low loss regime (up to around 65 km with APD or around 230 km with SNSPD). Furthermore, when there is a Trojan horse leakage with intensity $\nu = 10^{-3}$, we see that the decoy-state protocol (shown by black-dotted curves), suffers a more significant reduction in range as compared to the coherent state protocol (shown by green-dashed curves). This suggests that the coherent state protocol is more robust against Trojan horse attacks. We reiterate that in simulating the Trojan horse attack on decoy-state protocols, we assume that the Trojan horse attack does not affect the decoy-state implementation. Once we take into account the effect of Trojan horse attack on the decoy-state implementation, the key rate of the decoy-state protocol will suffer from an additional reduction due to the leakages.}
	\label{fig: decoy vs coherent keyrate}
\end{figure*}

After we obtain $Y^{\gamma,L}_{11}$ and $e^{\gamma,U}_{11}$, it is straightforward to apply our method or the quantum coin argument to lower bound the single-photon phase-error rate $\eph^{11}$. For decoy-state protocols, the lower bound on the asymptotic secret key rate is given by
\begin{equation}
    R \geq Q^Z_{11} [1 - h_2(\eph^{11})] - Q^Z_{\mu \mu} h_2(E^Z_{\mu \mu})
\end{equation}
where $Q^Z_{11} = \mu^2 e^{-2\mu} Y^Z_{11}$ is the single-photon gain in the $Z$ basis \cite{lo2012}.

Furthermore, we assume that Alice's and Bob's intensity modulators that choose the decoy settings have finite extinction ratio such that the intensity of the second decoy-state $\omega$ cannot be set to zero. In this case, we assume that the intensity of the second decoy-state is 1000 times smaller than that of the signal, i.e. $\omega = \mu / 1000$. Thus, for each point, the remaining free parameters in the optimization are the signal intensity $\mu$ and the first decoy-state intensity $\zeta$. The result of the simulation is given in Fig. \ref{fig: decoy-state MDI keyrate}.

Referring to Fig. \ref{fig: decoy-state MDI keyrate}, we again see an improvement in the lower bound on the secret key rate compared to the secret key rate obtained from the quantum coin technique. This confirms that our technique is able to give tighter secret key rate as compared to the quantum coin approach.

Although it is not an entirely fair comparison, one might be interested to compare the two protocols that we have analyzed. Fig. \ref{fig: decoy vs coherent keyrate} gives the superimposed plot of the key rate of the two protocols. When there are no leakages, as one would probably guess, the decoy-state protocol has a longer range than the coherent state protocol. However, rather surprisingly, we find that at short distances, the coherent state protocol gives a higher lower bound on the secret key rate than the decoy-state protocol.

A closer inspection reveals that for short distances, the coherent state protocol has significantly higher probability for successful Bell state measurements as compared to the single-photon gain of the decoy-state protocol. This is possible because the coherent state protocols rely on single-photon interference whereas decoy-state protocols are based on two-photon interference. Therefore, despite having higher phase-error rate, the coherent state protocol can give higher secret key rate than the decoy-state protocol at short distances. Nevertheless, at long distances, the fact that coherent state protocols have higher phase-error rate dominates and decoy-state protocols yield higher secret key rate in that regime. Drawing another connection to TF-QKD, single-photon interference is also an important feature of TF-QKD.

\begin{figure*}[t!] 
	\includegraphics[width=\textwidth, trim = {0 6.5cm 0 0}]{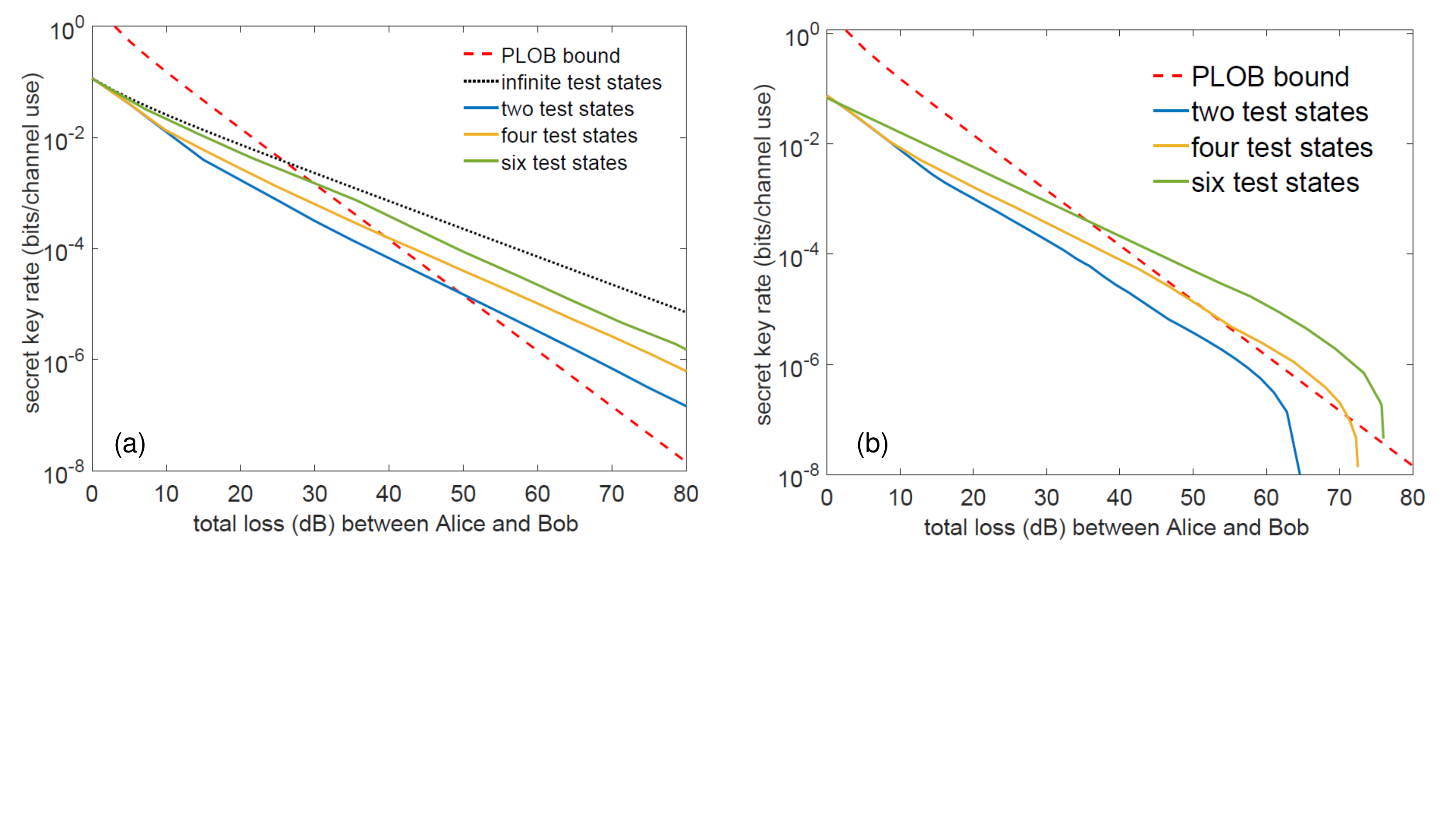} 
	\caption{\textbf{Key rate simulation for phase-matching MDI-QKD with finite test states for (a) loss-only scenario and (b) noisy scenario}. We compare the secret key rate obtained in phase-matching MDI-QKD with finite number of test states (represented by solid curves: from top to bottom, green curves for six test states, orange curves for four test states and blue curves for two test states) against the PLOB bound \cite{pirandola2017} (represented by red-dashed curves). For completeness, we also plot the secret key rate when infinite test states are used \cite{lin2018} (represented by black-dotted curve) in the noiseless scenario ($p_{dc} = 0$, $e_{\text{ali}} = 0$). Remarkably, in the noiseless scenario, two test states are sufficient to overcome the PLOB bound. Furthermore, when considering the lossless scenario, two test states give almost the same performance as infinite test states.}
	\label{fig: phase-matching MDI-QKD}
\end{figure*}

Interestingly, the coherent state protocols are more robust against Trojan horse attacks in the sense that the range of coherent state protocols suffers less penalty from Trojan horse attacks as compared to the range of decoy-state protocols. We also note that since we neglect the effect of leakages on the decoy-state implementation in the security analysis, the performance of the decoy-state protocol would suffer even more when we consider leakages in the intensity settings $\mu_A,\mu_B$.

\subsection{Phase-matching MDI-QKD with finite number of test states}
To demonstrate the versatility of our method, we will analyze a variant of the recently proposed twin-field QKD (TF-QKD) protocol called the phase-matching MDI-QKD \cite{lin2018}. In this protocol, for each round, Alice and Bob randomly select bases $x,y$ and bit values $a,b$. Then, they prepare the corresponding quantum states using the encoding given in equation \eqref{eq: phase-matching states} and send them to the central node for Bell state measurement. Note that Alice and Bob need to share a global phase reference, just like in the phase-encoding MDI-QKD protocol with non-phase-randomized coherent state. Consequently, to implement this protocol, phase-locking is also required.

As mentioned previously, there is a growing interest in TF-QKD.  Indeed, after the original proposal of TF-QKD protocol \cite{lucamarini2018}, other variants of the protocol have been proposed alongside with their respective security analyses \cite{ma2018, lin2018, tamaki2018information, curty2018simple, cui2019twin}. Remarkably, TF-QKD protocols are able to overcome the repeaterless bounds \cite{takeoka2014,pirandola2017}. However, many variants of the TF-QKD protocols (including the original proposal \cite{lucamarini2018}) require phase-randomization of two independent phase-locked lasers with the notable exception of Ref. \cite{lin2018}. There, the protocol uses non-phase-randomized coherent states which form phase-matching pairs. However, the results presented in Ref. \cite{lin2018} require Alice and Bob to send infinitely many coherent states, leaving the practical case where a finite number of test states is used as an open problem. Here, we show that a finite number of test states is indeed enough to overcome the repeaterless bound.

Remarkably, one can see phase-matching MDI-QKD as a generalization of the phase-encoding MDI-QKD that we have considered earlier in Section \ref{subsection: phase-encoding}. There, we assumed that the signal and test states share the same intensity. Here, we vary the intensity of different phase-matching pairs. More explicitly, in this protocol, Alice and Bob prepare states of the form
\begin{align}
    \ket{\As(a_x)} &= \ket{(-1)^a  e^{i \theta_x} \sqrt{\mu_x}} \nonumber\\
    \ket{\Bs(b_y)} &= \ket{(-1)^b e^{i \theta_y} \sqrt{\mu_y}} \label{eq: phase-matching states}
\end{align}
where $\theta_x, \theta_y$ are defined in similar way as in the phase-encoding MDI-QKD protocol and now the intensity of the coherent state depends on which basis Alice and Bob choose. By optimizing the intensity of each phase-matching pair, we show that we can overcome the repeaterless bound.

For comparison with the Ref. \cite{lin2018}'s result as well as the repeaterless bound, we will show the ideal performance of our protocol, i.e. when there are no dark counts or misalignment errors. For the repeaterless bound, we choose the secret key capacity $C$ given in Ref. \cite{pirandola2017}, also known as the Pirandola-Laurenza-Ottaviani-Banchi (PLOB) bound
\begin{equation}
    C = -\log_2(1-\eta) \label{eq: PLOB bound}
\end{equation}
where $\eta$ is the total transmittivity of the quantum channel. On the other hand, Ref. \cite{lin2018} proved that the secret key rate obtained when infinitely many test states are used in the loss-only scenario, denoted by $\bar{R}$, is given by
\begin{equation}
    \bar{R} = (1- e^{-2 \mu \sqrt{\eta}}) \left[1 - h_2\left( \frac{1-e^{-4\mu(1-\sqrt{\eta})}e^{-2\mu\sqrt{\eta}} }{2}\right ) \right] \label{eq: infinite test states}
\end{equation}

Lastly, it will also be interesting to see how the protocol fare in the presence of imperfections such as dark counts and misalignment errors. To that end, we will also use Parameter 2 of Table \ref{table: parameters} in our simulation. Furthermore, for fair comparison with PLOB bound, we plot the key rate against total loss (in dB), which includes the loss in the detector, instead of treating the loss in the fiber and in the detector separately. The plots of the key rate for the ideal and imperfect scenario as well as the PLOB and the key rate for the ideal infinite test states scenario are given in Fig. \ref{fig: phase-matching MDI-QKD}. Remarkably, two test states are sufficient to overcome the PLOB bound in the ideal scenario whereas in the noisy scenario, four test states are sufficient to overcome the PLOB bound. Furthermore, we emphasize that in our simulation, we use the states with the structure given by equation \eqref{eq: phase-matching states} and optimize (using a coarse-grained exhaustive search) the intensity of each test state. We leave the question of whether optimizing the global phases $\theta_x, \theta_y$ can significantly increase the secret key rate as an open problem.

\section{Conclusion}
To conclude, we have developed a new numerical technique to analyze the security of MDI-QKD protocols. This is done by employing SDP to obtain an almost-tight upper bound on the phase-error rate, hence the significant improvement in the distilled secret key rate compared to the quantum coin method. Additionally, our method also reveals that phase-encoding coherent state MDI-QKD might outperform decoy-state MDI-QKD at short distances. Furthermore, our method is highly efficient and can be applied to any protocols with discretely modulated, pure, input states including continuous-variable MDI-QKD protocols with discrete modulation \cite{ma2019long}. Due to the versatility of our technique, it can also be applied to MDI-QKD protocols which lack symmetry. For example, by applying our method to phase-matching MDI-QKD protocol, we also showed that the repeaterless bound can be overcome with only a few test states. It is also interesting to extend the results presented in this paper to allow mixed input states and to MDI-QKD networks where we have multiple transmitters as well as to include finite-key analysis.

\section*{Acknowledgements}
We thank Jie Lin and Norbert L\"utkenhaus for insightful discussions concerning the methods presented in Refs. \cite{coles2016, winick2018}. This research is supported by the National Research Foundation (NRF) Singapore, under its NRF Fellowship programme (NRFF11-2019-0001) and Quantum Engineering Programme (QEP-P2), the National University of Singapore (R-263-000-D04-731), and the Asian Office of Aerospace Research and Development (FA2386-18-1-4033).

\medskip

\bibliography{reference}

\onecolumngrid
\appendix
\newpage

\section{The phase-error rate of phase-encoding coherent state MDI-QKD} \label{section: Appendix phase error}
In the main text, we give a general procedure to obtain the expression for the phase-error rate for any MDI-QKD protocol. Here, we illustrate how this can be done. As a concrete example, we consider the phase-encoding coherent state MDI-QKD. In the virtual protocol, let the entangled states that Alice and Bob prepare for the key basis be denoted by
\begin{align}
    \ket{\Psi_\K}_{A'A} = \frac{1}{\sqrt{2}}
    \left[ \ket{+}_{A'} \ket{\sqrt{\mu}}_A + \ket{-}_{A'} \ket{-\sqrt{\mu}}_A \right] \nonumber\\
    \ket{\Psi'_\K}_{B'B} = \frac{1}{\sqrt{2}}
    \left[ \ket{+}_{B'} \ket{\sqrt{\mu}}_B + \ket{-}_{B'} \ket{-\sqrt{\mu}}_B \right]
\end{align}
It is easy to see that measuring the qubit systems in the $X$-basis is equivalent to sending the states $\{\ket{\sqrt{\mu}}, \ket{- \sqrt{\mu}} \}$ randomly. Now, the quantum channel is an isometry $\U$ that maps the system $AB$ to $EE'$ where the register $E$ stores Eve's quantum side information and the classical register $E'$ stores the result of the Bell state measurement. The states $\{\ket{\Psi^+}, \ket{\Psi^-}\}$ are associated to the respective Bell states, whereas the state $\ket{\varnothing}$ is associated to the case where the Bell state measurement fails to give a conclusive outcome. After the isometry, the global state is given by
\begin{align}
    \ket{\chi}_{A'B'EE'} = \frac{1}{2} \Bigg[& \ket{++}_{A'B'} 
    \left(\ket{\Es_{++}^{\Psi^+}}_E \ket{\Psi^+}_{E'} +
    \ket{\Es_{++}^{\Psi^-}}_E \ket{\Psi^-}_{E'} +
    \ket{\Es_{++}^{\varnothing}}_E \ket{\varnothing}_{E'} \right) \nonumber\\
    & + \ket{+-}_{A'B'} 
    \left(\ket{\Es_{+-}^{\Psi^+}}_E \ket{\Psi^+}_{E'} + \ket{\Es_{+-}^{\Psi^-}}_E \ket{\Psi^-}_{E'} + \ket{\Es_{+-}^{\varnothing}}_E \ket{\varnothing}_{E'} \right) \nonumber\\
    & + \ket{-+}_{A'B'} 
    \left(\ket{\Es_{-+}^{\Psi^+}}_E \ket{\Psi^+}_{E'} + \ket{\Es_{-+}^{\Psi^-}}_E \ket{\Psi^-}_{E'} + \ket{\Es_{-+}^{\varnothing}}_E \ket{\varnothing}_{E'} \right) \nonumber\\
    & + \ket{--}_{A'B'} 
    \left(\ket{\Es_{--}^{\Psi^+}}_E \ket{\Psi^+}_{E'} + \ket{\Es_{--}^{\Psi^-}}_E \ket{\Psi^-}_{E'} + \ket{\Es_{--}^{\varnothing}}_E \ket{\varnothing}_{E'} \right)
    \Bigg]
\end{align}
Post-selecting on the successful Bell state measurements,
\begin{align}
    \ket{\chi'}_{A'B'EE'} = \frac{1}{2\sqrt{P_\pass}} \Bigg[& \ket{++}_{A'B'} 
    \left(\ket{\Es_{++}^{\Psi^+}}_E \ket{\Psi^+}_{E'} +
    \ket{\Es_{++}^{\Psi^-}}_E \ket{\Psi^-}_{E'} \right) \nonumber\\
    & + \ket{+-}_{A'B'} 
    \left(\ket{\Es_{+-}^{\Psi^+}}_E \ket{\Psi^+}_{E'} +
    \ket{\Es_{+-}^{\Psi^-}}_E \ket{\Psi^-}_{E'} \right) \nonumber\\
    & + \ket{-+}_{A'B'} 
    \left(\ket{\Es_{-+}^{\Psi^+}}_E \ket{\Psi^+}_{E'} +
    \ket{\Es_{-+}^{\Psi^-}}_E \ket{\Psi^-}_{E'} \right) \nonumber\\
    & + \ket{--}_{A'B'} 
    \left(\ket{\Es_{--}^{\Psi^+}}_E \ket{\Psi^+}_{E'} +
    \ket{\Es_{--}^{\Psi^-}}_E \ket{\Psi^-}_{E'}\right)
    \Bigg]
\end{align}

Let the target state be $\ket{\Psi^+}_{A'B'}$. Conditioned on the outcome of the Bell state measurement, Bob applies the following correction on his qubit:
\begin{itemize}
    \item $ z = \Psi^+$: apply $\mathbb{I}_{B'}$
    \item $ z = \Psi^-$: apply $\Z_{B'}$
\end{itemize}
where $\mathbb{I}$ is the identity operator and $\{\X,\Y,\Z\}$ are the usual Pauli operators. After the corrections, the global state becomes
\begin{align}
    \ket{\chi''}_{A'B'EE'} = \frac{1}{2\sqrt{P_\pass}} \Bigg[& \ket{++}_{A'B'} 
    \left(\ket{\Es_{++}^{\Psi^+}}_E \ket{\Psi^+}_{E'} +
    \ket{\Es_{+-}^{\Psi^-}}_E\ket{\Psi^-}_{E'} \right) \nonumber\\
    & + \ket{+-}_{A'B'} 
    \left(\ket{\Es_{+-}^{\Psi^+}}_E \ket{\Psi^+}_{E'} +
    \ket{\Es_{++}^{\Psi^-}})_E \ket{\Psi^-}_{E'} \right) \nonumber\\
    & + \ket{-+}_{A'B'} 
    \left(\ket{\Es_{-+}^{\Psi^+}}_E \ket{\Psi^+}_{E'} +
    \ket{\Es_{--}^{\Psi^-}}_E \ket{\Psi^-}_{E'}\right) \nonumber\\
    & + \ket{--}_{A'B'} 
    \left(\ket{\Es_{--}^{\Psi^+}}_E \ket{\Psi^+}_{E'} +
    \ket{\Es_{-+}^{\Psi^-}}_E \ket{\Psi^-}_{E'} \right) \Bigg]
\end{align}

Since the target state is the state $\ket{\Psi^+}_{A'B'}$ and key is extracted by Alice and Bob by performing $\X$-measurement on each of their qubits, the phase-error rate can be computed by using
\begin{equation}
    \langle \Y_{A'} \otimes \Y_{B'} \rangle = 1 - 2 \eph
\end{equation}
where $\langle \Y_{A'} \otimes \Y_{B'} \rangle$ is simply
\begin{equation}
    \frac{1}{2 P_\pass} \Bigg( \real{ \braket{\Es_{++}^{\Psi^-}}{\Es_{--}^{\Psi^-}}  + \braket{\Es_{+-}^{\Psi^+}}{\Es_{-+}^{\Psi^+}} }
    -\real{ \braket{\Es_{++}^{\Psi^+}}{\Es_{--}^{\Psi^+}} +    \braket{\Es_{+-}^{\Psi^-}}{\Es_{-+}^{\Psi^-}}} \Bigg)
\end{equation}
and therefore, we have
\begin{equation}
    \eph = \frac{1}{2} + \frac{1}{4 P_\pass} \real{
    \braket{\Es_{++}^{\Psi^+}}{\Es_{--}^{\Psi^+}}
    -\braket{\Es_{++}^{\Psi^-}}{\Es_{--}^{\Psi^-}}
    -\braket{\Es_{+-}^{\Psi^+}}{\Es_{-+}^{\Psi^+}}
    +\braket{\Es_{+-}^{\Psi^-}}{\Es_{-+}^{\Psi^-}}}
\end{equation}
Thus, we obtain the phase-error rate of the phase-encoding coherent state MDI-QKD protocol. The phase-error rate of other protocols can be obtained through a similar procedure.

\section{Security analysis of MDI-QKD via the quantum coin technique} \label{section: quantum coin}
Here, we summarize the quantum coin technique presented in \cite{tamaki2012phase}. Consider a protocol where Alice (Bob) sends the state $\{ \ket{\As_0}_A, \ket{\As_1}_A \}$ ($\{ \ket{\Bs_0}_B, \ket{\Bs_1}_B \}$) to generate key and the state  $\{ \ket{\As_2}_A, \ket{\As_3}_A \}$ ($\{ \ket{\Bs_2}_B, \ket{\Bs_3}_B \}$) to test the channel. We define the following states
\begin{align}
    \ket{\Phi_x^A} &= \frac{1}{\sqrt{2}}
    \Big[ \ket{+}_{A'} \ket{\As_0}_A + \ket{-}_{A'} + \ket{\As_1}_A
    \Big] \nonumber\\
    \ket{\Phi_y^A} &= \frac{1}{\sqrt{2}}
    \Big[ \ket{+i}_{A'} \ket{\As_3}_A + \ket{-i}_{A'} + \ket{\As_2}_A
    \Big] \nonumber\\
    \ket{\Phi_x^B} &= \frac{1}{\sqrt{2}}
    \Big[ \ket{+}_{B'} \ket{\Bs_0}_B + \ket{-}_{B'} + \ket{\Bs_1}_B
    \Big] \nonumber\\
    \ket{\Phi_y^B} &= \frac{1}{\sqrt{2}}
    \Big[ \ket{+i}_{B'} \ket{\Bs_3}_B + \ket{-i}_{B'} + \ket{\Bs_2}_B
    \Big]
\end{align}

Now, consider a virtual protocol where Alice has an access to a quantum coin living in Hilbert space $\Co$. Furthermore, the entangled state describing the joint system of the coin, Alice's states and Bob's states, after taking into account sifting, is given by
\begin{equation}
    \ket{\Psi} = \frac{1}{\sqrt{2}} \Bigg[ \ket{0}_\Co  \ket{\Phi_x^A} \ket{\Phi_x^B} +
    \ket{1}_\Co  \ket{\Phi_y^A} \ket{\Phi_y^B}
    \Bigg]
\end{equation}
In this virtual protocol, the basis choice is made \textit{via} a $\Z$-measurement on the quantum coin $\Co$. It is sufficient to consider a single quantum coin system since the sifting step discards those events where Alice and Bob choose different bases.

The phase-error rate, $\eph$, is defined as the error-rate when Alice and Bob measure their the systems $A'B'$ in the $\Y$-basis, given that they send the states $\ket{\Phi_x^A}\ket{\Phi_x^B}$. However, the quantity that is measured in the actual protocol is the bit-error in the $\Y$-basis, denoted by $e_y$, which is defined as the error-rate when Alice and Bob measure the systems $A'B'$ in the $\Y$-basis, given that they send the states $\ket{\Phi_y^A}\ket{\Phi_y^B}$. Intuitively, if $\ket{\Phi_x^A}\ket{\Phi_x^B}$ is close to $\ket{\Phi_y^A}\ket{\Phi_y^B}$ (\textit{i.e.} when the basis-dependent flaw is small), $e_y$ will be close to $\eph$. This is quantitatively measured by the bias of the quantum coin when measured in the $\X$-basis. We define $\Delta$ as the probability of obtaining $\ket{-}_\Co$ when we measure the quantum coin in the $\X$-basis. Since the bias can be enhanced by the adversary by exploiting the failed Bell state measurements, the worst case scenario is when $\Delta$ is upper bounded by
\begin{equation}
    \Delta \leq \Delta_{\text{init}}/P_\pass
\end{equation}
where $\Delta_{\text{init}}$ is defined as
\begin{equation}
    \Delta_{\text{init}} =\frac{1 - \braket{\Phi_x^A}{\Phi_y^A} \braket{\Phi_x^B}{\Phi_y^B}}{2}
\end{equation}

By applying the so-called Bloch sphere bound \cite{tamaki2003unconditionally}, we have the following inequality
\begin{equation}
    1 - 2 \Delta \leq  \sqrt{e_y \eph} + \sqrt{(1-e_y) (1-\eph)}
\end{equation}
After some algebra, we can re-write the inequality such that we have an upper bound on $\eph$, we have
\begin{equation} \label{eq: QC_eph}
    \eph \leq e_y + 4 \Delta (1-\Delta)(1-2e_y) + 4(1-2\Delta)\sqrt{\Delta (1-\Delta) e_y (1-e_y)}
\end{equation}
Therefore, we can lower bound the secret key rate by plugging in equation \eqref{eq: QC_eph} to the following bound
\begin{equation}
    R \geq P_\pass \Big[ 1 - h_2(\eph) - h_2(e_x) \Big]
\end{equation}
\end{document}